# Synchronized clocks and time on a rotating disc

Richard Lenk

*Institute of Physics, Chemnitz University of Technology, 09107 Chemnitz, Germany.*
*Email: rlenk@physik.tu-chemnitz.de*

Abstract

Basic for the definition of "time" are clocks operating under stationary conditions. The periods of two clocks can be compared with each other via two return experiments. The central clock mediates between the rotating and the inertial frame. Dimensional arguments and a detailed deduction show that the period of a rotating clock, measured externally in the inertial system, is the same as the internally defined one. There is a common synchron-time, especially "absolute" simultaneity and invariance of flight times hold. This allows a new discussion of the twin problem.
The invariance of a rotating circle allows the transference of the inertial polar coordinates to the rotating plane. The flight time of a light pulse obeys a generalized Fermat's principle. If rays are well defined, this time of flight determines the phase shift of the wave and the result of interference experiments.
All general considerations and proofs are independent of the relativity theory, especially of the concept of local inertial systems and differential-geometrical techniques.

**Introduction**

As stated in a recent paper [1], after a century of discussion with a lot of controversies the subject of the physics on a rotating disc "is still alive". A large number of citations can be found there, additionally we quote [2]. In the present paper no discussion of all these contribution can be given. In our approach "time" is defined as a primary physical quantity for its own, realized by synchronized clocks. The method is
- independent of the special and/or general relativity theory
- specific for the rotational problem
- based on the invariant cycle numbers of clocks.

A conventional treatment employs the concept of local inertial systems, it is critically discussed in [2]. For our approach "inertial time" is also a prerequisite, but all general considerations are "global". The definition of "time" needs the solution of two independent problems: the comparison of periods, and a unique prescription for the simultaneity of distant events.
Both problems are solved for inertial systems, usually based on the established inertial symmetries: Clocks at different points in space the same period can be attributed to if they are identical as physical objects; and simultaneity is defined according to the Einstein-synchronisation. In some papers [3 and 4] this "gauge" is considered merely as the simplest choice, and anisotropic ones should be equally allowed. In the appendix 1, a possibility is sketched to exclude experimentally the claimed ambiguity. The inertial time is "isotropic" and univocally defined.

**Comparison of periods**

Different processes can be compared with each other only. These comparisons are standardized by clocks. The basic property of a clock is the existence of countable cycles. The numbers $\Delta N$ of cycles between two given "moments" are independent of the reference system and of the motion of clocks. Due to the presumend stationarity in time, the same time interval, the period T, can be attributed to each period of a given clock. The periods of two clocks at fixed positions can be compared directly. In two return experiments a signal is emitted from one of the clocks and reflected at the other clock back to the first one where the number $\Delta N_R(i|j)$ of cycles is counted. For light pulses there is no possibility to choose different signal velocities as for massive particles. Thus the path from 1 to 2 is univocally defined, and that from 2 to 1 too. In the case of rotation, these two paths are different. In both experiments the two pulses travel along these paths in different order only. Due to this equivalence the same time interval $T_1 \Delta N_R(1|2) = T_2 \Delta N_R(2|1)$ belongs to both return processes. The quotient of the periods is thus determined by that of the cycle numbers. The consistency relation $(T_1/T_2) \cdot (T_2/T_3) \cdot (T_3/T_1) = 1$ can be tested by three independent experiments. There is no need for a metrical analysis of the signal paths. The stationarity in time can be tested: The cycle numbers must remain the same if the experiment is repeated once and again.

The method allows to use different clocks in a given reference system, especially in an inertial system or in a frame rotating with constant angular velocity around a fixed axis, and in a static gravitational field, too. The periods of atomic clocks can be compared there, even combined with a common rotational motion as in GPS [5] if the gravitational potential acting on each of the clocks remains the same in the course of that motion. If it is not possible to transmit signal pulses directly, as on the surface of earth, a reflector can be placed between.

For a rotating frame all general problems are restricted to planes perpendicular to the axis. On a rotating disc the periods of all clocks C can be especially compared with the period of the central clock Z. For the internally determined period $T$(int) of a rotating clock one gets

$$T(\text{int}) \, \Delta N(CZC) = T(Z) \, \Delta N(ZCZ) = \Delta t(ZCZ) \, . \tag{1}$$

The second equality is valid because the central clock belongs to the rotating and the inertial system as well. Thus $\Delta t$ is the inertial time difference for the (ZCZ)-process, measured with the central clock.

The central clock belongs to the rotating and the inertial system as well. Thus the clock periods of both systems can be compared independently with the period of this common reference clock and, notwithstanding their relative motion, due to elementary logic also with each other. The mediation via the central clock is a unique aspect of the rotational problem. Methodically it is equally important as the relativity principle for translation. This is decisive for the allowing considerations.

**Period of moving clocks**

The return experiment can be also performed with two clocks moving commonly against an inertial system. For translation with constant velocity and for rotation with constant angular velocity, each of the two inertial signal paths is equivalent to the corresponding path at a later time, due to the inertial homogeneity in the first and to the isotropy in the second case. Thus the stationarity in time remains valid, and both moving clocks an externally, i.e. defined in the inertial system, period $T(\text{ext})$ can be attributed to, see below. The quotient of the invariant cycle numbers gives the quotient of these periods in the same manner as that of the $T(\text{int})$. Thus, for any clock, $T(\text{ext}) \sim T(\text{int})$ must hold, with a dimensionless factor that depends only on the parameters of the motion. For translation, $v/c$ is this parameter, whereas for rotation there is none. The translational time dilatation is therefore not in conflict with the invariance of $T_1/T_2$, whereas for rotation $T(\text{ext}) = T(\text{int})$ is the only possibility.

For a single moving clock C

$$T(\text{ext})\, \Delta N(C) = \Delta t_{12} \qquad (2)$$

is the definition of $T(\text{ext})$ where $\Delta t_{12}$ is the time difference between those two synchronized inertial clocks that the moving clock passes at the beginning and at the end of the measurement process not yet specified. For eq. (2) to be meaningful, however, the necessary homogeneity and isotropy, see above, must not be destroyed by the prescription for the setting of clocks. The Einstein-synchronization meets this demand.

For a rotating clock, the measurement process in eq. (2) can be specified to a return process via the central clock, $\Delta N(C) \rightarrow \Delta N(CZC)$. The comparison of eq. (2) with eq. (1) yields

$$\frac{T(\text{int})}{T(\text{ext})} = \frac{\Delta t(ZCZ)}{\Delta t(CZC)} \quad \rightarrow \quad 1. \qquad (3)$$

In the (ZCZ)-process, the moving clock acts only once (as "reflector"), whereas twice (as "source" and as "detector") in the (CZC)-process. As already discussed, due to the inertial isotropy the pulses travel along equivalent paths, and the two time differences in eq. (3) are equal. In accordance with the dimensional argument, $T(\text{int}) = T(\text{ext})$ holds. This statement can be tested by separate measurements in both reference systems.

If all the inertial clocks have the same period $T$, i.e. $\Delta t_{12} = T\, \Delta N_{12}$, and the rotating clocks are regulated via the central clock to achieve the same period, eq. (2) yields $\Delta N(C) = \Delta N_{12}$: Such a clock "ticks" exactly so as the inertial clocks passed. This is also true for a full circuit.

Generally, however, a regulated rotating clock is no more a twin of the inertial ones, and a non-regulated clock my have another period as its twin.

For the translational case, the period of a moving clock shows time dilatation TD as compared with the twin-clocks at rest. This general result follows from the relativity principle and holds if the twin criterion is correct. There is no problem with the latter for quantum objects. – For rotating clocks, especially for atomic clocks, TD is often assumed, based on the concept of local inertial systems. This may be true; there is no relationship to the invariance of the period in the sense of eq. (3).

**Common time**

With the invariance eq. (3) of the period, eq. (2) reads simply $\Delta t$(rot) = $\Delta t$(inert) for the time difference between two arbitrarily chosen events. If one of the rotating clocks is set, only once, to indicate the same time-value as the inertial clock just passed, this equality of times remains in the further course of motion: Each event the same time is attributed to in both reference systems. Especially, two events are simultaneous in the rotating system if they are so in the inertial one. The inertial consistency and univocality of the simultaneity of distant events is thus transferred to the rotating system. The setting of each clock described above amounts to an external prescription for the synchronization of the rotating clocks, see the next paragraph for the internal condition.

With common unit of time and "absolute" simultaneity, a common synchron-time is defined. The clocks of both systems realize this time; they constitute a single ensemble. This time concept is completely independent of a metric in space. "Common time" fits the success of GPS [5] where the synchron-time TAI on the rotating earth is matched with the time of the satellites rotating against distant stares.

In [2], the author postulates "absolute simultaneity" and combines it with differential-geometrical methods in the spirit of SRT. Relativistic time dilatation remains, however. In [6] absolute simultaneity is introduced to explain the Sagnac effect.

With common time, flight times of signal pulses can be determined in the inertial system for a rotating source-detector configuration and employed without change in the rotating system. If these times are measured, arbitrary signals are allowed. Are they to be calculated, the motion of the source must be considered. Generalizing the translational case, the independence of the inertial light propagation of that motion will be presumed for a rotating source, too. This statement is the essential "relativistic element" in our approach.

**Internal criterion of simultaneity**

Two light pulses are emitted simultaneously in different directions from two sources at $r_1$ and $r_2$. In the inertial system these pulses meet each other at all points with the same distance from $r_1$ and $r_2$, i.e. in two dimensions on the orthogonal mid-straight. At an arbitrarily chosen point of this line, the two pulses meet at a definite time. This defines a physical event that can, on principle, identified in any reference system. According to the common time, such a point defines especially the simultaneity on the rotating disc. This prescription is independent of a spatial metric.

For practical reasons, and in order to get quantitative results, it is useful to define in the inertial system a suitable registration line RL. In the course of its rigid rotation this line cuts the meeting straight at a special point that can be identified on the rotating plane. Its position can be described mathematically in a polar coordinate system, see the next paragraph. The simple results are already used in the following.

For a radial configuration the two sources are on the same rotating radial stick at distances $r_1$ and $r_2$ from the centre. This stick can be taken as RL, and geometry of plane triangles yields

$$R \cos\left(\frac{\omega}{c} [R^2 - r_1 r_2]^{1/2}\right) = \frac{1}{2}(r_1 + r_2) \tag{4}$$

for the position of the simultaneity point. The special case $r_1 = 0$ is sufficient for the setting of all clocks via the central one.

For an azimuthal configuration, the two sources define a rotating circle segment and the angle difference $\Delta\varphi$ between. The radial straight at $\Delta\varphi/2$ is the meeting line, and the segment ifself is a suitable RL. Each of the two pulses flights along its inertial secant, length l, and both meet the rotating RL after the flight time l/c at the position $\Delta\varphi/2 - \omega l/c$.

In the three-dimensional case, the inertial Einstein-criterion holds for all points of the axis, and each of these points is the centre of a rotating plane as discussed above – For two sources on a sphere, their positions define
   a) an "orthogonal" meeting plane that cuts the sphere on a non-rotating circle and
   b) the segment of a circle that connects the sources. This rotates about the given axis.
The common point of both circles defines simultaneity.

**Polar coordinates**

As shown in the first paragraph, the periods of rotating clocks can be compared directly with each other and, via the central clock, even with the periods of the inertial clocks. There is no equivalent possibility for the length of metric rods. Therefore, the definition of rotating lengths is to some extend a matter of choice. This is especially true for the often assumed and discussed [1 and 7] azimuthal length contraction if there is no independent definition of lengths in the rotating system. For "time" has been defined as independent of a spatial metric, the problem of length contraction is irrelevant in this context.

Times of flight will be calculated with polar coordinates r and $\varphi$ in the inertial system. To transfer these times to the rotating plane, it is necessary to define there coordinates $\tilde{r}$ and $\tilde{\varphi}$ in such a manner that no measurements of lengths on this plane are involved. This can be done in the simplest way. Without any markings, a rotating circle, a material ring, cannot be distinguished from one at rest. Its radius $r$ and its circumference $U$ can be measured directly, i.e. without clocks, with the inertial metric rods, and the results are obviously independent of its rotation. Thus $U = 2\pi r$, and the full plane angle is $2\pi$. All these circles constitute concentric marking lines that belong to the inertial and the rotating plane as well, and $\tilde{r} = r$ is the best choice, of course. Thus circular marking lines replace radial rotating rods.

Each of the circles can be divided in an arbitrary number n of equal segments. This can be done without measurements of lengths with the same method in the inertial or in the rotating plane. According to the rotational invariance of the circle, $\Delta\tilde{\varphi} = \Delta\varphi = 2\pi/n$ is the best definition of $\Delta\tilde{\varphi}$. To define $\varphi$ or $\tilde{\varphi}$ itself, a radial straight $\varphi = 0$ of $\tilde{\varphi} = 0$, respectively, is needed in the inertial or rotating plane to "synchronize" the division marks between the circle segments. The straight $\tilde{\varphi} = 0$ can be realized as the "projection" of a straight rotating in the inertial plane. The straightness of such a line can be tested by simultaneous registration of a sufficiently large number of points on it. This needs synchronized clocks, and the result depends on the setting of the clocks. This problem has been already discussed, and arguments for isotropic synchronization are given.

With the polar coordinates in the rotating plane defined, the coordinate transformation is given by the invariance of $r$ and by $\varphi \rightarrow \varphi - \omega t$ where $t$ is the common time. All relativistic effects are taken into account by the synchronization of the rotating clocks. Especially the free motion of a massive particle is "classical", influenced by Coriolis and centrifugal force.

**Radial configuration: time of flight, light-clocks and the twin problem**

Source and detector are fixed on the same rotating radial stick, one at $r_1$ and the other at $r_2$. The straight path that the light pulse travels along in the inertial systems is the third side in the triangle with sides $r_1$, $r_2$ and the angle $\omega l/c$ between them. If the source or the detector is at the centre, one gets $l = r$ for the length of the light path. For slow rotation

$$\omega r_i \ll c \quad : \quad l \approx \Delta r \left[1 - (\omega/c)^2 r_1 r_2\right]^{-1/2} \tag{5}$$

holds. This corresponds with the perpendicular configuration in the case of translation, the geometrical mean of the rotational velocities replaces v. The time of flight $l/c$ can be measured in the rotating system.

In a light clock, the pulse is reflected back and forth between two mirrors. The local character of such a clock needs a small mirror distance, $\Delta r \ll r_1 \approx r_2$. This justifies the same approximation as for slow rotation with the result

$$c\, T_L(r) = 2\, \Delta r M^{-1/2} \quad \text{with} \quad M \equiv 1 - (\omega r/c)^2 \; . \tag{6}$$

A regulated light clock with the smaller mirror distance $\Delta r = M^{1/2} \Delta r(Z)$ has the same period as the central clock Z.

A radial light-clock is a twin of the inertial light-clocks if the invariant mirror distance $\Delta r$ is the same. $2\Delta r/c = T$ is the period of the inertial twins, and $T_L(\text{twin}) = T M^{-1/2} > T$ is the period of the rotating twin which thus behaves in accordance with the so-called twin-paradox. This, however, is not at all strange because the rotating twin does not belong to the ensemble of clocks with the same period. According to the calculation above, $T_L(r)$ is first the externally, i.e. with the synchronized inertial clocks, measured period $T_L(\text{ext})$ of the rotating clock as defined in (2). This gives with $T(\text{ext}) \rightarrow T_L(\text{twin})$ and $\Delta N_C \rightarrow \Delta N_L(\text{twin})$, the relation $\Delta N_L = M^{1/2} \Delta N_{12}$ : the moving twin ticks more slowly than the inertial ones passed. The invariance $T_L(\text{ext}) = T_L(\text{int})$ yield the invariance of the time differences for any number of cycles; thus the concept of common time is reproduced.

For translational motion, $\omega r$ has to be replaced by v and $\Delta r$ by the perpendicular mirror distance. With these replacements, the relations for $T_L(\text{ext})$ and $\Delta N_L$ are the same as for the rotational case. There is full equivalence in this respect; a rotating light clock behaves in accordance with the concept of local inertial systems. There is, however, no way to compare directly the period of the moving clock with the clocks at rest. Instead of the invariance of periods, according to the relativity principle the same period $T$ is attributed to the moving twin in the co-moving inertial system as to the twins at rest. The invariant cycle number of the moving twin thus gives the time difference in the co-moving system if multiplied with $T$, and, if multiplied with $T_L(\text{twin})$, the time difference in that inertial system where the twin is moving. This is the time dilatation already mentioned.

Despite the equivalence of translation and rotation for $T_L$ and $\Delta N_L$, this result is combined with different theoretical concepts: {inequality $T(\text{int}) \rightarrow T \neq T(\text{ext})$ and time dilatation} for translation, but {equality and common time} for rotation. The symmetry formulated by the relativity principle is broken by the rotation.

**Time of flight for a general configuration**

The method is the same as already applied to the special radial configuration. Source S and detector D rotate commonly, on two circles with radii $r_1$ and $r_2$, but now with a non-zero angle difference $\Delta\varphi \equiv \varphi_D - \varphi_S$, counted as positive in the direction of rotation. The pulse travels along a straight, and its length $l$ is the third side in the triangle with sides $r_1$, $r_2$ and the angle $\Delta\varphi + \omega l/c$. For small values of $\Delta r = r_2 - r_1$ and $\Delta\varphi$, i.e. $\omega\Delta r \ll c$ and $|\Delta\varphi| \ll 1$, one gets with $\Delta l = c\Delta t$

$$cM(r)\Delta t = \lambda r^2 \Delta\varphi + [r^2(\Delta\varphi)^2 + M(r)(\Delta r)^2]^{1/2}$$
$$\text{with}\quad M(r) \equiv 1 - \lambda^2 r^2\ ,\quad \lambda \equiv \omega/c\ . \qquad (7)$$

For all physically allowed radii, $\lambda r < 1$ holds. The sign of the square root is always positive.

The linear term in (7), with $\Delta\varphi \gtrless 0$, reflects the inequivalence of pulse motion with or against rotation, often formulated as a difference of the velocity of light. This is the essence of the Sagnac effect, discussed in many papers, see [6] for a review, with different theoretical conceptions. Without terms of second order in $\omega r/c$ and generalized for the surface of the earth, the linear term is included in the definition of TAJ, the International Atomic Time.

With the calculated flight times the setting of clocks is experimentally simpler compared with the time-free (basic) method of meeting points. The light pulse between two clocks can transmit the information of the emission time.

The method explained above to define polar coordinates on the rotating disc is difficult to realize experimentally, and the idea of a "projection" can not be generalized to three dimensions. These problems can be overcome if the given flight times are, vice versa, used to define the coordinates. A circle of radius $r$, for example, is given by all points where light pulses emitted from the centre at time $t_0$ arrive at time $t_0 + r/c$. This method can be generalized, especially it applies to the three-dimensional case, too. This possibility underlines the importance of an independent time concept.

**Extremal principle**

The light pulse shall travel along an arbitrarily chosen curve $\varphi = \varphi(r)$ from one point 1 to a second point 2, both fixed on the rotating plane. The corresponding time $\Delta t_{12}$ can be calculated by integration along this curve,

$$c\, \Delta t_{12} = \int_1^2 dr\, M^{-1}(r) \left\{ \lambda r^2 \frac{d\varphi}{dr} \pm \left[ r^2 \left(\frac{d\varphi}{dr}\right)^2 + M(r) \right]^{1/2} \right\} \quad \text{for} \quad dr \gtrless 0. \tag{8}$$

The real path, under the condition of free propagation, is determined by the principle of shortest travelling time. $\varphi$ is a cyclic variable, and the variation yields the differential equation

$$\lambda r^2 \pm Q^{-1} r^2 \frac{d\varphi}{dr} = \text{const} \cdot M(r) \quad \text{for} \quad dr \gtrless 0 \tag{9}$$

where $Q$ is the positive square root in (8). The integration constant is determined by the end points.

In the inertial system, the real path is a straight line. The transformation $\varphi \to \varphi - \omega t$ gives the curved path wanted. If the function $\varphi(r)$ thus constructed is a solution of the differential equation, the real path obeys the extremal principle. In the very simple special case $r_1 = 0$ or $r_2 = 0$, the inertial path is a radial straight outwards or inwards, thus $\varphi(\text{inert}) = \text{const}$ and $dr = \pm c\, dt$. This gives $d\varphi/dr = \mp\lambda$ and $Q = 1$ in the rotating system, and the differential equation is fulfilled with const = 0. The general proof is sketched in appendix 2. It supports the concept of common time.

**Remarks on frequency and phase of light waves**

Despite of spatial inhomogeneity and anisotropy, the frequency *f* of light, emitted by a source at rest in the rotating system, should be the same everywhere. Equivalent to this statement, there is no Doppler effect in the inertial system for a common rotation of source and detector.

Like the period of a rotating clock, the frequency of a source or a resonant absorber is generally not the same as those of the inertial "twins". These frequencies are to be measured. Resonant absorption in Mößbauer experiments [8] should be perturbed if source and absorber are at different distances of the centre. For a central source, the frequency is not changed by the rotation.

For short wave lengths, a bundle of rays can be attributed to the wave field. For a point-source S, a flight time $\Delta t$(PS) is defined along the ray from S to any point P. The extremal property of $\Delta t$ renders sensible the thesis that the field at P, as the result of a *constructive* interference of elementary waves (in the sense of the Huygens-Fresnel principle), oscillates with the phase difference $2\pi f \, \Delta t$(PS), thus $u(t) \sim \exp[2\pi f(t-\Delta t_{PS})]$. With this representation of the wave field, inhomogeneity and anisotropy of the propagation process are taken into account compactly by the time of flight. For example, the condition of periodicity for a resonator reads $f_n[\Delta t(12) + \Delta t(21)] = n$. For a ring one gets $f_n T_\pm = n$ where $T_\pm$ is the flight time around, either with or against the rotation. If relativistic corrections of order $(\omega r/c)^2$ can be neglected, the linear term in (7) yields the contribution $\pm 2\omega A/c$ where A is the area surrounded. This gives, generalized to a sphere, the frequency of a ring laser, especially the difference $f_+ - f_-$. The measurement of this difference [9 and 10] yields a precise sensor for the earth rotation.

For two paths from the source to the detector the difference of the flight times determines the intensity in the interference experiment. A suitable parameter can be adjusted to keep maximum intensity for varying angular velocity.

**Appendix 1: Inertial isotropy of light propagation and test of simultaneity criteria**

Time differences, especially times of flight, obey (trivially) cyclic zero-relations for three or more events. These relations remain true if to the $\Delta t(ij)$ terms $\Delta(ij)$ are added that depend only on the positions of the events in a projective manner, compare the Lorentz transformation. This yields anisotropy and changes the definition of simultaneity. Sums of flight times, as in ring processes or in coincidence experiments with arbitrary reflex points between source and detector, remain unchanged in the first example or get a *constant* additive in the second one. Thus anisotropy cannot be excluded that way.

Differences of flight times determine the lines where two pulses, emitted flash-like from two sources, meet each other. For all points of such a line, $\Delta t(ij)$ has the same value. In the case of isotropy, these lines are hyperbolas with the sources as focal points. Each non-trivial change of the flight times yields other geometrical forms. Thus anisotropy could be detected without clocks and times.

The analysis of meeting lines needs the definition of coordinates in space. A corresponding metric can be useful, it is, however, not necessary. The information coded in the meeting lines should be sufficient to determine flight times univocally, and thus the simultaneity of distant events, too. This statement is not restricted to inertial systems.

**Appendix 2: Proof of the extremal principle**

For a straight line in the inertial system, the equation $r(\varphi) = A/\cos\varphi$ holds. $A = \min r$ defines the radial ray with $\varphi = 0$. For $d\varphi > 0$, the pulse moves "with the rotation". $x = r \sin\varphi$ is the co-ordinate along the straight, thus $dx/dt = +c$. From the above relation one gets the geometrical relations

$$W \cdot r \left(\frac{d\varphi}{dr}\right)_0 = \pm \frac{A}{r}, \qquad W \cdot \frac{dx}{dr} = \pm 1 \qquad \text{for} \quad dr \gtrless 0 \tag{10}$$

with $W(r) \equiv [1 - (A/r)^2]^{1/2}$.

For the path in the rotating plane follows

$$r \frac{d\varphi}{dr} = r \left[\left(\frac{d\varphi}{dr}\right)_0 - \omega \frac{dt}{dr}\right] = \pm \frac{A/r - \lambda r}{W(r)} \qquad \text{for} \quad dr \gtrless 0 \ . \tag{11}$$

Now the square root $Q$ defined in the text after (9) can be evaluated analytically with the result $Q = W^{-1}(1-\lambda A)$, and the differential equation is fulfilled with const $= A[1-\lambda A]^{-1}$, $\lambda A < 1$.

For $d\varphi < 0$, the sign of the $(A/r)$-term is changed, and the integration constant takes the value $-A[1+\lambda A]^{-1}$.